\documentclass[aps,twocolumn]{revtex4}
\usepackage{amsmath, amsthm,times}
\usepackage{epsfig}
\usepackage{amssymb}
\usepackage{graphicx}

\begin{document}

\sloppy

\title{Splitting the wavefunctions of two particles in two boxes}

\author{S.J. van Enk}

\address{
Department of Physics, University of Oregon\\
Oregon Center for Optics and
Institute for Theoretical Science\\
 Eugene, OR 97403}
\begin{abstract}
I consider two identical quantum particles in two boxes. We can split each box, and thereby the wavefunction of each particle, into two parts.  When two half boxes are interchanged and combined  with the other halves, where do the two particles end up? I solve this problem for two identical bosons and for two identical fermions. The solution can be used to define a measurement that yields some information about the relative phase between the two parts of a split wavefunction.
\end{abstract}
\maketitle
\section{Introduction}
Consider a particle confined to a 1-D box in the groundstate. Suppose we  split the box in two equal parts, in such a way that the probability of finding the particle in each half is 50\%. Quantum mechanics ascribes a nonzero wavefunction to each half. If one believes the particle must really be in one of the two half boxes, then it is hard to say what it means to ascribe a wavefunction to the half box where the particle is, in reality, {\em not}. According to Einstein, who took this realistic stance, this argument  shows quantum mechanics cannot be complete, and in that sense is a precursor to the more famous EPR paradox \cite{EPR}. 
In Ref.~\cite{norsen} philosophical implications of  ``Einstein's boxes'' are discussed in great detail, while Ref.~\cite{gea} analyzes how one can actually split the wavefunction of a particle in a box by slowly raising a potential barrier.

Here I consider a variation of this problem involving two boxes containing one particle each (see Figures 1 and 2). For convenience, let us call one box red and the other blue.
We split each of the two colored boxes in two equal halves, a left half and a right half. Suppose we combine (without measuring where the particles are) the blue left half with the red right half; and we combine the blue right half with the red left half, where combining is  the inverse process of splitting. The question is: what is the probability of finding both particles in the blue/red box? Let us denote this probability by $P_{br}$. By symmetry it should equal the probability $P_{rb}$ to find both particles in the red/blue box.
For classical distinguishable particles the answer is obvious: there is a 25\% chance we find two particles in the blue/red box, 25 \% chance we find two particles in the red/blue box, and 50\% chance that we find 1 particle in each. Hence classically we have $P_{br}=P_{rb}=1/4$.

But now consider the  case of identical quantum particles. Where do the particles end up with what probabilities when we have initially two identical bosons in identical spin states? And what happens when we have initially two identical fermions in identical spin states? 
\begin{figure}
  \includegraphics[width=9cm]{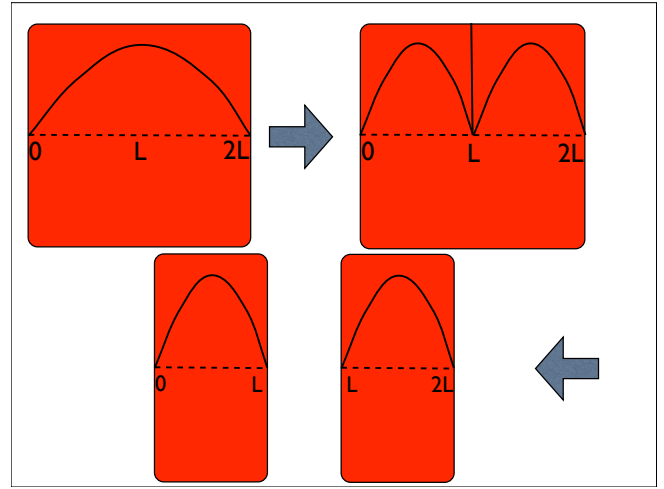}
  \caption{Splitting of the wavefunction: we start with a particle in the ground state of a 1-D box of length $2L$.  We split the box, and thereby the wavefunction, by slowly raising a symmetric potential barrier in the center. This yields two boxes with length $L$ and a 50\% probability to find the particle in either half.}
\end{figure} 
\begin{figure}
  \includegraphics[width=9cm]{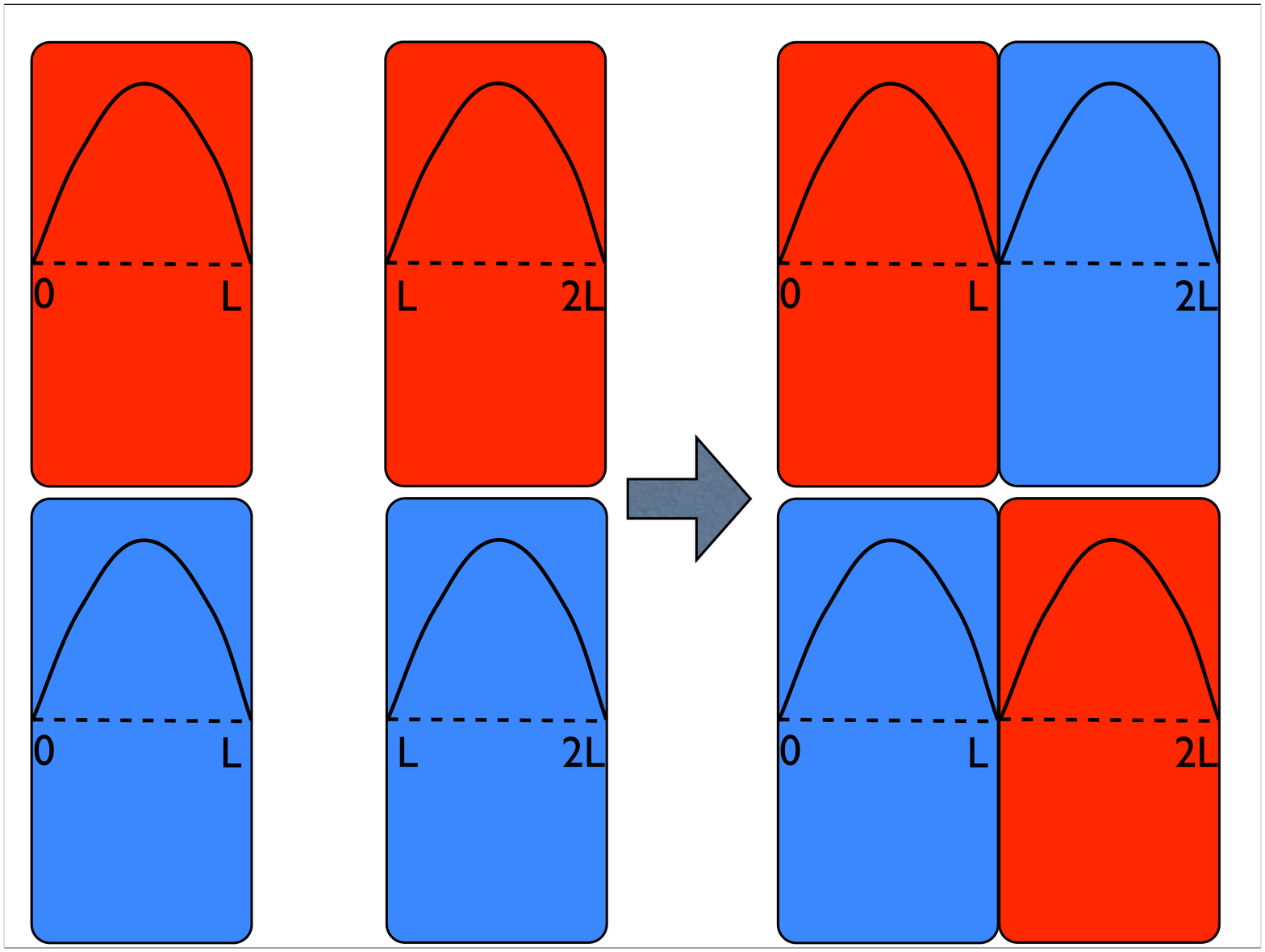}
  \caption{Recombining the boxes: we split the wavefunction of two boxes, a red one and a blue one. Then we recombine the two halves, but only after interchanging the red and blue half boxes on the right. 
Finally, we remove the potential barrier in the middle (not shown). What is the probability to find the particles together in the upper red/blue box?}
\end{figure} 
There seem to be four plausible \footnote{In an informal poll among 14 scientists present at a meeting at the Perimeter Institute, Waterloo, Canada, in June 2008, the voting distribution for answers 1--4 was: 2,2,4,2, respectively, with 4 choosing the fifth answer ``I don't know.''} answers (I give the answers in boxes, but I also provide some pseudo-arguments, which should not be taken too seriously, in favor of each) 
\begin{enumerate}
\item The classical answer still holds, because we consider only one kind of observable, particle number, so there are no quantum effects arising from noncommuting observables.
Hence \fbox{$P_{br}=P_{rb}=1/4$}.
\item Bosons like to huddle together, and they always end up together in 1 box (and the effect is similar to the Hong-Ou-Mandel effect in optics \cite{HOM}, where two identical photons impinging on the two different input ports of a 50/50 beamsplitter always end up together at one of the two output ports). Hence, \fbox{$P_{br}=P_{rb}=1/2$ for bosons}. Fermions in the same spin state, on the other hand, avoid each other, and so \fbox{$P_{br}=P_{rb}=0$ for fermions}.
\item Since all we are doing is exchanging identical particles, the final situation is not different in essence from the initial state, apart from the different coloring of the boxes. Thus each particle should end up in 1 box again, all by itself. Thus, \fbox{$P_{br}=P_{rb}=0$} for both bosons and fermions.
\item Argument 3 is correct for bosons, hence \fbox{$P_{br}=P_{rb}=0$ for bosons}. But exchanging two identical fermions leads to an extra minus sign in the wavefunction. This extra minus sign turns destructive interference into constructive interference, and {\em vice versa}, and hence \fbox{$P_{br}=P_{rb}=1/2$ for fermions}. 
 \end{enumerate}
 In order to answer this question we could use second-quantized quantum mechanics. This is not necessary, however, and for simplicity I will stick to the language of wavefunctions. After all, that is how the question is phrased.
\section{Where do the particles end up?}
We write the initial two-particle wavefunction as a function of two coordinates $x_{1,2}$,  
\begin{equation}
\Psi(x_1,x_2)={\cal S}[\psi_{rg}(x_1)\psi_{bg}(x_2)],
\end{equation}
where the subscripts $r$ and $b$ refer to the red and blue boxes, $g$ stands for ground state, and ${\cal S}$ stands for the symmetrization operator (for bosons) or antisymmetrization (for fermions) operator. 
The splitting of the wavefunction is represented as
\begin{eqnarray}
\Psi(x_1,x_2)&\rightarrow& \frac{1}{2}{\cal S}\left[(\psi_{rL}(x_1)+\psi_{rR}(x_1))(\psi_{bL}(x_2)+\psi_{bR}(x_2))\right]\nonumber\\
&:=&\Psi'(x_1,x_2),
\end{eqnarray}
where $L$ and $R$ stand for left and right, respectively.
Writing out this product gives four terms,
\begin{eqnarray}\label{box4}
\Psi'(x_1,x_2)&=& \frac{1}{2}{\cal S}[
\psi_{rL}(x_1)\psi_{bL}(x_2)\nonumber\\
&&+\psi_{rL}(x_1)\psi_{bR}(x_2)\nonumber\\
&&+\psi_{bL}(x_1)\psi_{rR}(x_2)\nonumber\\
&&+\psi_{rR}(x_1)\psi_{bR}(x_2)]
\end{eqnarray}
This form of the wavefunction is sufficient to see that the probability for both particles to end up in the red/blue box is $P_{rb}=1/4$: it is only one term (the second term) out of four equal-amplitude terms that contributes to this probability. This is true for both bosons and fermions. Thus answer 1 is correct.  Where, one might wonder at this point, does the bosonic or fermionic character of the particles manifest itself?

An important ingredient of the complete answer is the fact that the first excited state of the particle in the full-sized box will be split in a very similar way as the ground state \cite{comment}. Namely, there are two degenerate ground states for the two half boxes, differing only in the sign (phase) of the wavefunction of the right half box relative to that of the left half box.
That is, we have the invertible (unitary, in fact) mapping (see Figure 3)
 \begin{eqnarray}\label{split} 
 \psi_{g}(x)\leftrightarrow \psi_+:=[\psi_L(x)+ \psi_R(x)]/\sqrt{2}\nonumber\\
  \psi_{e}(x)\leftrightarrow \psi_-:=[\psi_L(x)- \psi_R(x)]/\sqrt{2}
\end{eqnarray}
\begin{figure}
  \includegraphics[width=9cm]{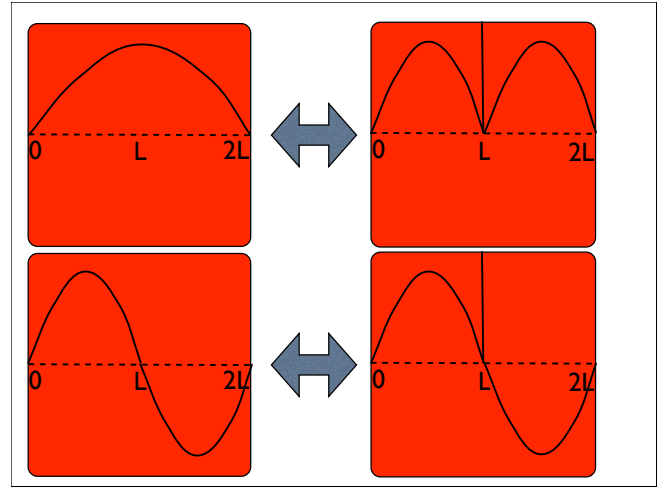}
  \caption{The ground-state wavefunction and the first excited-state wavefunction are split in similar ways, ending up in degenerate ground states of the split boxes. We can write the wavefunction-splitting process and its inverse as
  $\psi_{g,e}(x)\leftrightarrow [\psi_L(x)\pm \psi_R(x)]/\sqrt{2}$.}
\end{figure}
Consider now the second term of (\ref{box4}) in more detail. In order to see what the merging of the two boxes accomplishes, we note that
$\psi_{rL}$ is an equal superposition of $\psi_+$ and $\psi_-$ with a zero relative phase, and $\psi_{bR}$ is an equal superposition of the same two states but with a $\pi$ relative phase. Thus we have 
\begin{eqnarray}
{\cal S}[\psi_{rL}(x_1)\psi_{bL}(x_2)]&=&
\frac{1}{2}{\cal S}[\psi_+(x_1)\psi_+(x_2)]\nonumber\\
&-&\frac{1}{2}{\cal S}[\psi_+(x_1)\psi_-(x_2)]\nonumber\\
&+&\frac{1}{2}{\cal S}[\psi_-(x_1)\psi_+(x_2)]\nonumber\\
&-&\frac{1}{2}{\cal S}[\psi_-(x_1)\psi_-(x_2)].
\end{eqnarray}
The inverse transformation acting on just this term, describing both particles in the red/blue box, gives 
\begin{eqnarray}\label{final}
{\cal S}[\psi_{rL}(x_1)\psi_{bL}(x_2)]&=&
\frac{1}{2}{\cal S}[\psi_g(x_1)\psi_g(x_2)]\nonumber\\
&-&\frac{1}{2}{\cal S}[\psi_g(x_1)\psi_e(x_2)]\nonumber\\
&+&\frac{1}{2}{\cal S}[\psi_e(x_1)\psi_g(x_2)]\nonumber\\
&-&\frac{1}{2}{\cal S}[\psi_e(x_1)\psi_e(x_2)].
\end{eqnarray}
\subsubsection{Bosons}
Taking into account the symmetrization of the wavefunction, we see that for bosons the second and third term in (\ref{final}) cancel.
Thus, the two bosons, if they end up in the red/blue box, are either both in the ground state, or both in the first excited state. That is where the bosonic character of the particles manifests itself, not in the probability of finding the two bosons in one box. This effect is related to the above-mentioned Hong-Ou-Mandel effect \cite{HOM}. 
\subsubsection{Fermions}
Taking into account the antisymmetrization of the wavefunction, we see that for fermions the first and fourth term in (\ref{final}) are identically zero.
Thus, the two fermions, if they end up in the red/blue box, must be in different states, one in the ground state, one in the first excited state. That is where the fermionic character of the particles manifests itself, not in the probability of finding the two fermions in one box.
\subsubsection{Discussion}
Let us consider here the pseudo arguments in favor of the various wrong answers given in the Introduction.
The argument for answer 2 is incorrect: there is no principle that says that fermions cannot be in the same box, only that they cannot be in the same state. 
The argument for answer 3 is incorrect, because we do not interchange two identical particles, we only interchange two boxes in which one may or may not find a particle. The argument for answer 4 is incorrect for the same reason, although there is indeed an extra minus sign for fermions under the exchange $rR\leftrightarrow bR$: it appears in only one term, namely in the wavefunction 
\begin{equation}
{\cal S}[\psi_{rR}(x_1)\psi_{bR}(x_2)]=-{\cal S}[\psi_{bR}(x_1)\psi_{rR}(x_2)].
\end{equation}
Let us also comment on the correct answer.
The probabilities to find the particles in their half boxes do not change {\em before} the half boxes are merged. But the merging itself is a local operation, in that there cannot be any population density current flowing from one pair of half boxes in one location to the other pair in a different location. Thus the merging operation also does not change the probabilities. Hence the probabilities $P_{br}$ and $P_{rb}$ must be the same as those obtained classically.

We can use our example to note the distinction between ``identical'' and ``indistinguishable'' particles. In the initial situation we have two identical particles, but they are distinguishable: one is in a red box, the other in a blue box. It is only in the final situation in the bosonic case that we may end up with  two identical {\em and} indistinguishable particles, namely when they end up in the same box {\em and} in the same state. 

Similarly, it is only in the final situation that we can observe quantum interference. Consider, e.g., the state where two particles end up inside one box in different states, one in the ground state, the other the excited state. There are two ways the two particles can end up in that situation: the particle in the ground state could originate from the blue box or from the red box. For identical bosons these two pathways interfere destructively [and hence the two bosons cannot end up in different states], for two identical fermions they interfere constructively.
\section{Distinguishing $\psi_+$ from $\psi_-$}
Suppose we split the wavefunction of a particle in a box, and take the two halves far apart. We thus obtain the highly delocalized wavefunction $\psi_+(x)=[\psi_L(x)+\psi_R(x)]/\sqrt{2}$. From this wavefunction we can produce the orthogonal wavefunction $\psi_-(x)=[\psi_L(x)-\psi_R(x)]/\sqrt{2}$  if we apply a $\pi$ phase shift 
to the wavefunction for, say, the right box (applying the phase shift to the left box gives the same result up to an irrelevant overall minus sign). That is, we apply a small time-dependent perturbation $V(t)$ (independent of $x$) for a finite amount of time  such that $\int dt V(t)/\hbar=\pi$. 
\begin{figure}
  \includegraphics[width=9cm]{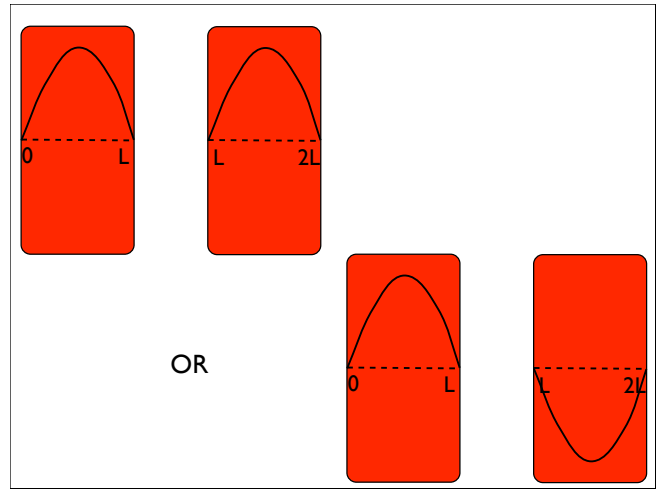}
  \caption{Can we distinguish these two split wavefunctions? The two possible wavefunctions assigned to the right-hand side box differ only in sign and so are locally the same physically.}
\end{figure}

Consider the following scenario (see Fig.~4): someone hands us a delocalized split box, promises us  either $\psi_+$ or $\psi_-$ was prepared, each with probability 1/2, but does not tell us which one of the two. Can we distinguish the two orthogonal states described by $\psi_+$ from $\psi_-$? {\em Locally}, in each location the wavefunctions inside the box are the same up to an irrelevant overall phase factor. Nevertheless, if we bring the two boxes together again, we can certainly distinguish the two states by  performing a state measurement on the particle. The measurement outcomes of ground state and excited state correspond to $\psi_+$ and $\psi_-$, respectively. 
Suppose, however, we are not allowed to bring the two boxes close together.  Can we distinguish the two states?

By using the results from the preceding Section we find we can indeed distinguish the two states, at least with 50\% probability, as follows: we  first create a second box containing an identical particle, and split it into two halves in a {\em known} state, say $\psi_+$.  We then locally merge the two halves on each side. Rewrite the joint states using
\begin{eqnarray}\label{same}
{\cal S}[(\psi_L(x_1)+ \psi_R(x_1))(\psi_L(x_2)+\psi_R(x_2))]=\nonumber\\
\sqrt{2}{\cal S}[\psi_+(x_1)\psi_+(x_2)-\psi_-(x_1)\psi_-(x_2)]\nonumber\\
+{\cal S}[\psi_L(x_1)\psi_L(x_2)]+
{\cal S}[\psi_R(x_1)\psi_R(x_2)]
\end{eqnarray}
and
\begin{eqnarray}\label{diff}
{\cal S}[\psi_L(x_1)- \psi_R(x_1))(\psi_L(x_2)+\psi_R(x_2))]=\nonumber\\
\sqrt{2}{\cal S}[\psi_+(x_1)\psi_-(x_2)-\psi_-(x_1)\psi_+(x_2)]\nonumber\\
+{\cal S}[\psi_L(x_1)\psi_L(x_2)]+
{\cal S}[\psi_R(x_1)\psi_R(x_2)]
\end{eqnarray}
for the $+$ and $-$ states, respectively.
The only  distinction between these two states is the first term, which describes a state of one particle on each side with the particles either  in the {\em same} state [Eq.~(\ref{same})], or in {\em different} states [Eq.~(\ref{diff})].  
So, when we detect one particle on each side [which occurs with 50\% probability] we project onto those two different (orthogonal) states. By Eq. (\ref{split}) merging the two half boxes converts $\psi_+$  into the ground state and $\psi_-$  into the first excited state. We thus simply perform state measurements after merging the boxes on each side: If the two measurement outcomes are the same (different), then we must have been given $\psi_+$ ($\psi_-$). 

For this measurement the bosonic or fermionic character is irrelevant, and the measurement works equally well for both types of particles. A similar problem and a similar solution were also considered in \cite{vaidman}.  This solution  is a simple (partial) solution for the case where we are not allowed to create or destroy particles. 

There are better solutions (with larger success probabilities) if we allow particle creation and annihilation and describe the problem in second-quantized form. In particular, if we allow measurements 
on each box that project onto states $|\pm\rangle:=[|0\rangle \pm |1\rangle]/\sqrt{2}$, where $|0\rangle$ and $|1\rangle$ are states with  0 and 1 particles in the ground state, then that measurement perfectly (with 100\% success probability) distinguishes the two states $\psi_+$ and $\psi_-$. The reason is that we can write these two states in second-quantized form as
\begin{equation}
\psi_{\pm}\leftrightarrow
[|1\rangle_L|0\rangle_R\pm |0\rangle_L|1\rangle_R]/\sqrt{2},
\end{equation}
and in turn rewrite these second-quantized states by using
\begin{equation}
|1\rangle_L|0\rangle_R+ |0\rangle_L|1\rangle_R
=|+\rangle_L|+\rangle_R- |-\rangle_L|-\rangle_R,
\end{equation}
and
\begin{equation}
|1\rangle_L|0\rangle_R- |0\rangle_L|1\rangle_R
=|+\rangle_L|-\rangle_R- |-\rangle_L|+\rangle_R.
\end{equation}
That is, if the  results of the above-mentioned projective measurements on the two boxes are the same (different), then we must have been given $\psi_+$ ($\psi_-$).
This solution, too, is valid both for bosons and fermions.

Disallowing particle-number nonconserving measurements boils down to imposing a super-selection rule. For further reading on super-selection rules in the context of quantum information protocols, see the review article Ref.~\cite{review} and references therein. 
\section{Conclusion}
The problem of merging two boxes containing two identical quantum-mechanical particles makes the standard problem of identical  particles in a box a bit more challenging and, hopefully, more interesting. 
It also allows one to contemplate the meaning of the relative phase between different parts of the wavefunction of a delocalized particle in a simple context.
Finally, the problem of finding a measurement of that phase provides a simple example of the second-quantization formalism yielding a superior solution to the approach based on wavefunctions only.


\begin{thebibliography}{99}

\bibitem{EPR}A. Einstein, B. Podolski, and N. Rosen, {\em Can quantum mechanical description of physical reality be considered complete?}, Phys. Rev. {\bf 47}, 777--780 (1935).

\bibitem{norsen}T. Norsen,	{\em Einstein's boxes}, Am. J. Phys. {\bf 73}, 164--176 (2004); see also
A. Shimony, {\em Comment on Norsen's defense of Einstein's ``box argument''}, Am. J. Phys,. {\bf 73}, 177-178 (2005).

\bibitem{gea}J. Gea-Banacloche, {\em Splitting the wavefunction of a particle in a box}, Am. J. Phys. {\bf 70}, 307--312 (2002).

\bibitem{HOM}Hong, C. K. and Ou, Z. Y. and Mandel, L.,
{\em Measurement of subpicosecond time intervals between two photons by interference},
Phys. Rev. Lett.,
{\bf 59},
2044--2046,
(1987).

\bibitem{comment}M. Lakner and J. Peternelj, {\em Comment on ``Splitting the wavefunction of a particle in a box''},Am. J. Phys. {\bf 71}, 519 (2003).




\bibitem{vaidman}Aharonov, Y.  and Vaidman, L., {\em Nonlocal aspects of a quantum wave},
Phys. Rev. A
  {\bf 61},
  052108-1--11 (2000).
  
\bibitem{review}
S.D. Bartlett, T. Rudolph, and R.W. Spekkens, {\em  Reference frames, superselection rules, and quantum information}, Rev. Mod. Phys. 79, 555-610 (2007).


\end{thebibliography}
\end{document}